# Establishing the carrier scattering phase diagram for ZrNiSn-based half-Heusler thermoelectric materials

Qingyong Ren[1], Chenguang Fu[2], Qinyi Qiu[3], Shengnan Dai[4], Zheyuan Liu[1], Takatsugu Masuda[5], Shinichiro Asai[5], Masato Hagihala[6], Sanghyun Lee[6], Shuki Torri[6], Takashi Kamiyama[6,7], Lunhua He[8,9,10], Xin Tong[10,11], Claudia Felser[2], David J. Singh[12], Tiejun Zhu[3], Jiong Yang[4], and Jie Ma[1,13]

[1]Key Laboratory of Artificial Structures and Quantum Control, School of Physics and Astronomy, Shanghai Jiao Tong University, 800 Dongchuan Road, Shanghai 200240, China
[2]Max Planck Institute for Chemical Physics of Solids, Nöthnitzer Straße 40, 01187 Dresden, Germany
[3]State Key Laboratory of Silicon Materials, School of Materials Science and Engineering, Zhejiang University, Hangzhou 310027, China
[4]Materials Genome Institute, Shanghai University, 99 Shangda Road, Shanghai 200444, China
[5]Neutron Science Laboratory, Institute for Solid State Physics, University of Tokyo, Kashiwanoha, Kashiwa, 277-8581, Japan
[6]Institute of Materials Structure Science, High Energy Accelerator Research Organization (KEK), Tokai, Ibaraki 319-1106, Japan
[7]Department of Materials Structure Science, Sokendai (The Graduate University for Advanced Studies), Tokai, Ibaraki 319-1106, Japan
[8]Beijing National Laboratory for Condensed Matter Physics, Institute of Physics, Chinese Academy of Sciences, Beijing 100190, China
[9]Songshan Lake Materials Laboratory, Dongguan, Guangdong 523808, China
[10]Spallation Neutron Source Science Center, Dongguan 523803, China
[11]Institute of High Energy Physics, Chinese Academy of Sciences, Beijing 100049, China
[12]Department of Chemistry and Department of Physics and Astronomy, University of Missouri-Columbia, Columbia, MO 65211, USA
[13]Shenyang National Laboratory for Materials Science, Institute of Metal Research, Chinese Academy of Sciences, Shenyang, 110016, China

Correspondence and requests for materials should be addressed to C.G.F. (email: Chenguang.Fu@cpfs.mpg.de) or to J.Y. (email: jiongy@t.shu.edu.cn) or to J.M. (email: jma3@sjtu.edu.cn).

**Abstract**

Chemical doping is one of the most important strategies for tuning electrical properties of semiconductors, particularly thermoelectric materials. Generally, the main role of chemical doping lies in optimizing the carrier concentration, but there can potentially be other important effects. Here, we show that chemical doping plays multiple roles for both electron and phonon transport properties in half-Heusler thermoelectric materials. With ZrNiSn-based half-Heusler materials as an example, we use high-quality single and polycrystalline crystals, various probes, including electrical transport measurements, inelastic neutron scattering measurement, and first-principles calculations, to investigate the underlying electron-phonon interaction. We find that chemical doping brings strong screening effects to ionized impurities, grain boundary, and polar optical phonon scattering, but has negligible influence on lattice thermal conductivity. Furthermore, it is possible to establish a carrier scattering phase diagram, which can be used to select reasonable strategies for optimization of the thermoelectric performance.

## 1. Introduction

Thermoelectric (TE) materials enable direct conversion between heat and electricity. They can be used to generate electricity based on the Seebeck effect when a thermal gradient exists or to transfer heat against temperature gradient based on the Peltier effect when an electric current is applied.[1] Many thermoelectric materials are being explored for power generation applications, such as half-Heusler,[2] PbTe,[3, 4, 5] silicides,[6] CoSb$_3$,[7] and Mg$_3$Sb$_2$.[8] The energy conversion efficiency for TE materials depends heavily on the TE material performance. Practical applications require further improvement on the performance of TE materials. This is quantified by the thermoelectric figure of merit, $zT = \alpha^2 \sigma T/(\kappa_{ele} + \kappa_{lat})$, where $\alpha$, $\sigma$, $T$, $\kappa_{ele}$, and $\kappa_{lat}$ are the Seebeck coefficient, the electrical conductivity, the absolute temperature, the electronic thermal conductivity, and the lattice thermal conductivity, respectively.[9] Good TE materials generally have features as Phonon Glass Electron Crystal, which means good electrical transport properties, characteristic of crystalline semiconductors, and a low thermal conductivity, characteristic of glasses.[10]

The optimization of TE performance is complex, requiring manipulation and optimization of both the electrical and thermal transport.[11] This includes controlling the underlying scattering mechanisms, *e.g.* alloy scattering,[12] ionized impurity scattering,[13] carrier-carrier scattering,[14] acoustic and optical phonon scatterings[15] for the charge carriers, and phonon anharmonicity,[16, 17, 18] atomic rattling,[19] nanostructuring,[20] glass-like or superionic behavior[21, 22, 23] for the phonons. Due to the diverse temperature dependences of the scattering mechanisms, electrical and thermal conductivities, in general, exhibit different temperature dependences. For instance, the carrier mobility follows $\mu_{ac} \propto T^{-3/2}$ for acoustic phonon dominated scattering, $\mu_{pe} \propto T^{-1/2}$ for piezoelectric scattering, or $\mu_i \propto T^{3/2}$ for ionized impurity scattering,[24] while the lattice thermal conductivity follows $\kappa_U \propto T^{-1}$ for Umklapp scattering or $\kappa_{al} \propto T^{-1/2}$ for alloy scattering.[25, 26]

On the other hand, the scattering rates for the carrier scattering mechanisms also show strong carrier concentration dependences (*n* for electron concentration, *p* for hole concentration), and may be more complex than the standard parabolic band expressions noted above. As an example, changes in carrier concentration could alter the carrier energy or carrier density of states around the Fermi level, and hence influence the acoustic phonon scattering and the mobility.[27] As demonstrated in Figure 1(a), the carrier transport in most TE semiconductors, except the lead chalcogenides, are dominated by the phonon scattering, and the mobility shows almost monotonous decrease with the carrier concentration, *n* or *p*.[15, 28, 29, 30, 31] The temperature and carrier concentration dependences of the scattering provide possibilities for the manipulation of the transport properties. Therefore, mapping of a carrier scattering phase diagram in the space of temperature and carrier concentration would provide a useful guidance for optimization or design of TE materials, such as the study of BXGa (X = Be, Mg, and Ga) compounds.[32]

Half-Heusler (HH) TE compounds, which are used for promising power generation applications, have attracted extensive attentions in recent years.[33, 34] Since the study of ZrNiSn compounds,[35, 36, 37] large *zT* values above unity can be obtained in both *n*-type and *p*-type HH alloys.[38, 39] The carrier transport mechanism underlying their high TE performance has been a focus of study. Current studies of the carrier transport mechanisms in half-Heusler compounds mainly focus on acoustic phonons, alloy, and grain boundary scatterings. These exhibit distinct temperature dependences.[12, 40, 41, 42] On the other hand, the carrier concentration dependent carrier mobility, $\mu(n / p)$, of HH compounds has been rarely studied. As summarized in Figure 1(b), $\mu(n / p)$ of both *n*-type and *p*-type HH compounds exhibits a non-monotonous behavior and the peak $\mu(n / p)$ usually occurs near to the optimal $n / p$ of the system, where the *zT* peaks.[40, 41, 42, 43] This indicates that the carrier scattering mechanisms in HH compounds have strong dependences on the carrier concentration. The decrease of $\mu(n / p)$ at high $n / p$ region is generally thought to be a result of the acoustic phonon scattering, as exhibited in Figure 1(a). However, the increase of $\mu(n / p)$ in the low $n /$

*p* region suggests some other important scattering mechanisms, which are not well understood and motivate further study.

Here, we use high-quality single and polycrystalline samples, and various probes, including electrical and thermal transport measurements, neutron powder diffraction (NPD) and inelastic neutron scattering (INS) measurements, and first-principles calculations, to identify the electron and phonon transport behavior of Sb-doped ZrNiSn, a representative HH TE system. We reveal important scattering beyond the acoustic phonon scattering. Specifically, ionized impurity, grain boundary, and polar optical phonon scatterings are shown to contribute significantly in controlling the electrical transport. Moreover, the effect of electron-phonon interactions on lattice thermal conductivity is comprehensively analyzed. Based on these results and analysis, a carrier scattering phase diagram is established. This could serve as a guide in further improvement of the TE performance in HH compounds.

## 2. Results

**Electrical and thermal transport properties**

High-quality ZrNiSn$_{1-x}$Sb$_x$ polycrystalline samples were made by levitation melting and spark plasma sintering (see Methods for details). All the polycrystalline samples contain 5-7 % more interstitial Ni on the *4d* (¾, ¾, ¾) site (Supplementary Figure 1), being consistent with the previous electron probe microanalysis.[40] Temperature dependences of the mobility in these polycrystalline samples were studied over the temperature range of 5 K to 315 K (variation of $n_H$ with $T$ and Sb content are shown in Supplementary Figure 3). $\mu(T)$ of Sb-doped samples monotonically decreases with temperature (Figure 2(a)), since the acoustic phonon and alloy scatterings contribute importantly to the electron scattering. Meanwhile, the undoped ZrNiSn parent sample exhibits a positive dependence of $\mu(T)$ on temperature, opposite to the doped samples. Similar to *n*-type Mg$_3$(Sb,Bi)$_2$ system, this kind of increasing trend of $\mu(T)$ raises the issue of the carrier scattering mechanism in underdoped TE materials with low carrier concentration or with small grain size,[44, 45] *i.e.*, does the increased $\mu(T)$ below the crossover temperature stems from the grain boundary scattering or ionized impurity scattering?[44, 45]

One direct way to answer this question would be a comparative study of the temperature dependent electrical conductivity of both polycrystalline and single-crystalline samples, thereby addressing the issue of grain boundary scattering. High-quality single crystalline ZrNiSn$_{1-x}$Sb$_x$ samples were then grown by a Sn-flux method[46] (details are given in Methods). The corresponding $\mu(T)$ is presented in Figure 2(b). The $\mu(T)$ of the undoped ZrNiSn single crystal, which by nature does not have grain boundaries, still shows an increasing trend with $T$. This indicates that ionized impurities might dominate the electron transport. The $\mu(T)$ for the Sb-doped ZrNiSn$_{1-x}$Sb$_x$ crystals exhibits a decreasing trend with $T$, similar to the doped polycrystalline samples (Figure 2(a)). The carrier concentration dependent $\mu(n)$ at 300 K of both polycrystalline and single-crystalline samples is displayed in Figure 2(c). The $\mu(n)$ of single crystals is larger than that of polycrystalline samples at low *n* region (< $10^{21}$ cm$^{-3}$) while they converge at the high *n* heavily doped region (> $10^{21}$ cm$^{-3}$), indicating that grain boundaries also play a significant role in scattering electrons of lightly doped polycrystalline samples. Additionally, the non-monotonous dependence of $\mu(n)$, as observed in the representative polycrystalline HH systems (Figure 1(b)), is also shown in the single crystalline ZrNiSn$_{1-x}$Sb$_x$ samples.

We now turn to the thermal transport. The lattice thermal conductivity $\kappa_{lat}$ of polycrystalline ZrNiSn$_{1-x}$Sb$_x$ samples over the temperature range of 2 K to 300 K is shown in Figure 2(d) (temperature dependences of the electronic resistivity $\rho$, Lorenz number $L$, electronic thermal conductivity $\kappa_{ele}$, and total thermal conductivity $\kappa_{tot}$ are provided in Supplementary Figure 4). It is surprisingly found that over the whole studied temperature range Sb doping does not show a significant effect on the $\kappa_{lat}$ of ZrNiSn. This is true even for the extreme case where 12% of Sn was replaced by aliovalent Sb. The atomic mass and radius of Sb are close to those of Sn. Nonetheless, the fact that such a high content of aliovalent doping does not obviously affect the phonon propagation is unexpected, since in many solid materials even partial isotopic substitution can lead to a significant changes in the $\kappa_{lat}$.[47, 48] In the very similar case of Sn-doped ZrCoSb$_{1-x}$Sn$_x$ HH compounds, replacement of 10% of the Sb by Sn results in a large reduction of ~45% in the $\kappa_{lat}$ near room temperature.[43] Similarly, in p-type NbFeSb-based compounds, aliovalent doping also significantly suppresses the lattice thermal conductivity.[33, 49] Therefore, the present result together with the non-monotonous dependence of $\mu(n)$ motivate further exploration of the underlying electron and phonon transport mechanisms in the ZrNiSn$_{1-x}$Sb$_x$ system.

**Phonon density of states**

To understand the effects of electron doping on the electron and phonon transport properties in ZrNiSn-based HH compounds, it is necessary to consider electron-phonon interactions. We begin with the phonons. Phonon density of states (DOSs) measurements for ZrNiSn$_{1-x}$Sb$_x$ with different electron concentrations were carried out on the High-Resolution Chopper Spectrometer, BL12 HRC, at MFL of J-PARC. The momentum and energy dependence of powder-averaged dynamical structure factor, $S(\mathbf{Q}, E)$, of ZrNiSn at 300 K, as an example, are shown in Figure 3(a). The phonon signal shows normal $\mathbf{Q}^2$ dependence. The neutron-weighted phonon DOSs were obtained by integration of the $S(\mathbf{Q}, E)$ over the range of momentum transfers $0.4 \leq \mathbf{Q} \leq 6.0$ Å$^{-1}$. The carrier concentration dependence of neutron-weighted phonon DOSs of ZrNiSn$_{1-x}$Sb$_x$ is shown in Figure 3(b). Three phonon bands, 0-18.5 meV, 18.5-25.5 meV, and 25.5-33 meV, are observed, corresponding to one acoustic phonon band and two optical phonon bands, respectively. Due to the different ratios of neutron scattering cross-section over atomic mass, $\sigma/m$, of the different elements, the motions of Ni are overemphasized while those for Zr and Sn/Sb are underestimated. The phonon DOS as obtained from first-principles calculations shown in Figure 3(c) for ZrNiSn with proper neutron weighting is in good agreement with experimental data (see Supplementary Figure 5 for calculated phonon DOSs without neutron weighting).

**Electron-phonon interaction & lattice thermal conductivity**

A careful comparison of the phonon DOSs in Figure 3b shows that the optical phonon bands exhibit clear changes with $n$. A prominent tail is present above 30 meV for the parent ZrNiSn sample, which gradually becomes smaller with increasing $n$. However, the calculated phonon transport properties of ZrNiSn demonstrate that the optical phonons above ~25 meV have much smaller group velocity (Figure 3(d)), and suffer stronger scattering (Figure 3(e)). Hence, they have a quite small contribution to the average lattice thermal conductivity, $\kappa_{lat}$, (less than ~1.5%, Figure 3(f)), and the changes in the optical phonons cannot make a palpable influence on $\kappa_{lat}$. In addition, the acoustic phonon band in Figure 3(b) is insensitive to the changes in $n$. This is one indication of relatively weak coupling between electrons and acoustic phonons. This weak coupling could be attributed to the symmetry-protected non-bonding orbitals-electron states or impotent acoustic phonons deformation potentials.[50] Since the acoustic phonons are the main carriers for thermal energy with large group velocity (Figure 3(d)) and small scattering rate (Figure 3(e)), it might be the overall weak interaction between the electrons and heat-carrying acoustic phonons and the negligible contribution of the optical phonons (above ~25 meV) that make the $\kappa_{lat}$ does not show obvious changes in the polycrystalline ZrNiSn$_{1-x}$Sb$_x$ compounds with different carrier concentrations or Sb contents (Figure 2(d)).

**LO-TO splitting and polar optical phonon scattering**

We turn to the observed non-monotonous behavior of $\mu(n)$ (Figures 1(b) and 2(c)). The 25.5-33 meV optical phonon band of the ZrNiSn parent sample, which contains the prominent tail, can be decomposed into two peaks as shown in Figure 4(a). To understand the origin of the tail, first-principles calculations of phonon dispersion are performed. As shown in Figure 4(b), the degeneracy of longitudinal optical (LO) phonon and transverse optical (TO) phonon at ~28 meV is lifted around the Brillouin zone center, corresponding to the LO-TO splitting. The peaks centered at ~28.9 meV in Figure 4(a) corresponds to the TO branches, while the ~31.9 meV one represents the LO branch.

The LO-TO frequency (energy) difference is due to the long-range electric field in an insulator.[51] This polarization field is produced by the LO phonon vibration, which in turn can raise the frequency of LO branch (see methods for details). In addition, this polarization field also imposes a strong disturbance on the carrier transport, *via* the Fröhlich electron-phonon interaction.[27, 52] This coupling between electrons and polar optical phonons can be assessed by the dimensionless polar coupling constant:

$$\alpha_{\text{PO}} = \frac{e^2}{4\pi\hbar} \left(\frac{m^*}{2\hbar\omega_{\text{LO}}}\right)^{1/2} \left(\frac{1}{\varepsilon_\infty} - \frac{1}{\varepsilon_s}\right) \quad (1)$$

$\hbar$ is the Planck constant, $m^*$ is the effective mass of electrons, $\omega_{\text{LO}}$ is the LO frequency, and $\varepsilon_\infty$ and $\varepsilon_s$ are the high-frequency and static dielectric functions. The observation of the LO-TO splitting in ZrNiSn confirms that HH compounds exhibit polar chemical bonds.[53] This mixed bonding can also be important for the thermal conductivity.[54] The polar coupling constant, $\alpha_{\text{PO}}$, for ZrNiSn is estimated as 0.32, indicating that the polar optical phonon scattering could play an important role in electrical transport properties.

We now turn to the collapse of the tail in the phonon DOSs with Sb doping (Figure 3(b)). The optical phonon energy is connected to the reduced-atomic-mass by the relation: $\omega_{\text{op}} \propto M_{\text{r}}^{-1/2}$, where the reduced-atomic-mass is defined as $M_{\text{r}} = 1/\sum_a (1/M_a)$, where $M_a$ is the mass of atom $a$ in unit cell. However, the modification of the optical phonon energy by 12% doping of Sb is less than 0.04%, which is too small to explain the change observed in the phonon DOSs.

Instead, the screening effect needs to be considered to explain the collapse of the LO-TO splitting. The presence of free carriers and the concomitant screening can modify both the electron-LO phonon interaction and the dispersion of the LO optical phonons, such as Kohn anomaly of the phonon branch in metals.[55] For the LO-TO splitting, the additional restoring force of the LO vibrations is a Coulomb interaction. This Coulomb interaction can be well screened by free carriers. This can be described in the Thomas-Fermi approximation (see methods for details).[56] Higher free carrier concentration leads to better screening, shorter Thomas-Fermi screening length, $r_{\text{TF}}$, and hence smaller Coulomb interaction and LO-TO splitting as schematized in Figure 4(c). The phonon DOSs (in Figure 4(d)) from first-principles calculations with consideration of the screening effect reproduce this change (detailed changes in the phonon dispersion with Sb doping are given in Supplementary Figure 6). This change is also confirmed by the reduced LO-TO splitting as determined from the phonon DOSs measurements (see Figure 4(e)). This reduction of LO-TO splitting could also be well simulated (dashed line in Figure 4(e)) with the screening model as described with Equations (2) and (4) in Methods. In addition, this reduced LO-TO splitting or polarization field is accompanied by a raising mobility, $\mu_{\text{po}}(n)$, that is limited by polar optical phonon scattering.[40] However, the release of the polar optical phonon scattering is accompanied by an enhancement of the acoustic phonon scattering as shown in Figure 2(c). This crossover lead to an optimal carrier concentration and mobility around $n = 3\text{-}5 \times 10^{20}$ cm$^{-3}$ for the TE performance of ZrNiSn-based compounds (Figure 2(c)).

Following the screening scenario, additional free charge carriers could enhance the screening effect in all semiconductors, and transitions between different carrier scattering mechanisms are expected from the low $n$ / $p$ region to the high $n$ / $p$ region, hence a non-monotonous carrier mobility. However, of the semiconductors listed in Figure 1, only the lead chalcogenides and half-Heusler compounds are known to exhibit non-monotonous behavior at room temperature. To explain the different dependences of carrier mobility on carrier concentration, the polar coupling constants of the half-Heusler compounds are calculated using the Equation (1) (see Supplementary Table 1 and Supplementary Figure 7 for details). A comparison of the polar coupling constants between different samples is shown in Figure 4(f). Compared with the lead chalcogenides and half-Heusler compounds, the polar coupling is much weaker in InSb, $CoSb_3$, and GaAs compounds, indicating smaller, and perhaps negligible polar optical phonon scattering in those compounds. Therefore, screening of the polar optical phonon scattering does not make important changes to the carrier mobility in those compounds. In contrast, the screening of polar optical phonon scattering in the more strongly-polar lead chalcogenides and half-Heusler compounds would noticeably tune the dominant carrier scattering mechanisms, leading to the non-monotonous behavior.

Finally, we comment on the details of the phonon modes based on first-principles calculations. The results are shown in the insets of Figure 4(b). The primitive unit cell of ZrNiSn contains three atoms, and therefore, there are 6 optical phonon branches. The three high energy optical phonon branches correspond to the LO-TO splitting, while the other three low energy optical phonon branches do not present obvious LO-TO splitting. For both the high and low energy branches, the Ni and Sn atoms vibrate in opposite directions. This is consistent with the molecular orbitals analysis of ZrNiSn, where the Ni and Sn atoms primarily form $(NiSn)^{4-}$ clusters and then bond with the $Zr^{4+}$ cations to form the ZrNiSn compound.[57] The difference between the high and low energy branches relates to the vibration of Zr. Given that the Ni atoms sit at the centers of both the Zr and Sn tetrahedrons, the opposite vibration of both the Zr and Sn with respect to the Ni would involve relatively larger chemical bonding forces, and hence correspond to the higher energy optical phonon branches. The small LO-TO splitting for the low energy optical phonon can then be attributed to the relatively larger group velocity and small displacement of the $(NiSn)^{4-}$ cluster with respect to the Zr sublattice.

## 3. Discussion

The effects of dielectric screening on carrier mobility are not just limited to the polar optical phonon scattering.[58, 59] Rather we find other important effects. Polar optical phonon scattering becomes weak at low temperature, because of the Bose factor that means that there are no excited optical phonons at low temperature. However, the non-monotonous behavior of the $\mu(n/p)$ does not disappear at low temperature, but becomes more evident as depicted in Supplementary Figure 8. This phenomenon is attributed to the screening of the ionized impurity and grain boundary scatterings, since these are the dominant scattering mechanisms at low temperatures. Taken the ionized impurity scattering as an example, the potential of the ionized impurities in heavily doped semiconductor cannot be described by the bare Coulomb potential, $\varphi(r) \propto \frac{1}{r}$, any more, while a new formula in the Thomas-Fermi approximation, $\varphi(r) \propto \frac{1}{r}\exp(-\frac{r}{r_{TF}})$, should be applied.[56] As schematized in Supplementary Figure 9, the screened potential drops more quickly than the bare Coulomb potential, and then the scattering is weaker.

According to the definition of the Thomas-Fermi screening length (see Equation (5) in Methods), the screening effect becomes weaker with increasing temperature. Nonetheless, it is still effective at higher temperature, even at 800 K.[40] This is important for the manipulation of the carrier scattering mechanisms in the ZrNiSn-based compounds. In addition, our previous study demonstrated that the polar optical phonon, acoustic phonon, and alloy scatterings, which are the main carrier scattering mechanisms at 300 K, still play the dominated roles at higher temperature, and crossover between these scattering mechanisms can also been realized by introducing the screening effect.[40]

In summary, different carrier scattering mechanisms in ZrNiSn-based half-Heusler compounds are studied with high-quality single and polycrystalline crystals, combined with electrical transport measurements, neutron powder diffraction and inelastic neutron scattering measurements, as well as theoretical calculations. With increasing carrier concentration $n$, a crossover of the carrier mobility $\mu(n)$ and optimal TE performance are found in the region of $n = 3–5 \times 10^{20}$ cm$^{-3}$. This crossover of $\mu(n)$ corresponds to the variation of the dominant scattering mechanisms, from ionized impurity and grain boundary scattering to acoustic phonon/alloy scatterings at low temperature, or from polar optical phonon scattering to acoustic phonon/alloy scatterings at high temperature. These crossovers between different dominant scattering mechanisms are realized through the screening of the ionized impurity, grain boundary and polar optical phonon scattering, and the enhancement of acoustic phonon/alloy scatterings. On the bases of these insights into the different carrier scatterings, a carrier scattering phase diagram may be derived. This is shown in Figure 5, and could be used as a guide for the further improvement of the TE properties in the half-Heusler TE materials. For example, grain boundaries, which are frequently employed to reduce lattice thermal conductivity, have an adverse effect on carrier transport in lightly doped half-Heusler compounds, e.g. NbFe$_{0.95}$Ti$_{0.05}$Sb.[60] This would work against the efforts to suppress the lattice thermal conductivity in improving TE performance. However, the grain boundary scattering for carrier transport could be well screened in heavily doped semiconductor, *e.g.* NbFe$_{0.8}$Ti$_{0.2}$Sb, and the introduction of grain boundaries could successfully improve the TE performance.[49] We expect the establishment of such a carrier scattering phase diagram in other TE systems could also help to choose reasonable and effective strategies for the optimization of TE performance.

### 4. Methods

**Synthesis:** Ingots of nominal ZrNiSn$_{1-x}$Sb$_x$ ($x$ = 0.00, 0.04, 0.08 and 0.12) compounds were synthesized by levitation melting of stoichiometric amount of Zr, Ni, Sn and Sb under argon atmosphere. The melting was carried out three times to ensure homogeneity. The ingots were ground into powder, and then sintered using spark plasma sintering (SPS-1050, Sumitomo Coal Mining Co.) at 1193 K for 10 min under a pressure of 65 MPa in vacuum. The densities of the as-sintered samples are 95% of theoretical densities. Then, the SPS-samples were used for both physical properties and neutron scattering measurements. NPD measurements (Supplementary Figure 1) do not show discernible impurity phases. All samples show extra Ni at the vacancy site, and the occupancy varies between 4.9% and 6.4%, being consistent with previous results from Electron Probe Microanalysis measurements.[40]

The undoped and Sb-doped ZrNiSn$_{1-y}$Sb$_y$ single crystals were grown by a Sn Flux method. The starting powders of Zr, Ni, Sn, and Sb were mixed together in a molar ratio of 1 : 1 : 10 : $y$ ( $y$ = 0 ~ 0.10). The mixtures were sealed in a dry quartz tube under high vacuum, which was heated up to 1100 °C in 15 h and further dwelled for 24 h. After that, the tube was slowly cooled down to 650 °C at a rate of 2 °C/h. Single crystals were obtained after a centrifuging process to remove the Sn flux. The as-grown single crystals with typical size of 2 mm were checked and oriented at room temperature by a Laue X-ray diffractometer, which show sharp and distinct diffraction spots (Supplementary Figure 2), indicating high quality and crystallinity.

**Physical properties:** The electrical and thermal transport properties are characterized using the Physical Property Measurement System (PPMS) with the electrical transport option and thermal transport options. To keep consistent, the electrical resistivity was measured along [110] direction for all the single crystals, while the magnetic field was applied along [1$\bar{1}$1] direction during the Hall resistivity measurement. Hall carrier concentration $n_H$ was calculated using the equation $n_H = 1/(eR_H)$, where $e$ is the unit charge and $R_H$ is the Hall coefficient. The carrier mobility $\mu_H$ was calculated using $\mu_H = R_H/\rho$. The Lorenz number is calculated by the single parabolic band model.[12]

**Neutron powder diffraction:** ~3 grams samples were ground into powder in agate mortar and used for NPD measurements. The measurements were taken on the Super High Resolution Powder Diffractometer, BL08 SuperHRPD, at the Material and Life Science Experimental Facility (MLF) of Japan Proton Accelerator Research Complex (J-PARC) and on the General Purpose Powder Diffractometer, GPPD, at the China Spallation Neutron Source. Vanadium sample can were used in the measurements. The NPD patterns were collected over the temperature range of ~10 K to ~300 K. Rietveld refinement were performed using the Z-Rietveld software.

**Inelastic neutron scattering:** The inelastic neutron scattering measurements were performed using the High Resolution Chopper Spectrometer, BL12 HRC, at MFL of J-PARC. ~13 grams of powder samples were wrapped in aluminium foil and then sealed in thin-walled cylindrical Al can, filled with low-pressure helium gas. The incident neutron energies, $E_i = 66$ meV, was used with an energy resolution (full width at half maximum) of $\Delta E/E_i$ ~5%. The measurements were performed at 10 K, 100 K, 200 K and 300 K with a closed-cycle He refrigerator. Correspondingly, the empty Al can was also measured in identical conditions at all temperatures. All of the spectra were normalized with respect to the total incident flux. The time-of-flight data were reduced with the HANA (a program for data reduction collected on HRC). Then, the obtained powder-averaged dynamical structure factor $S(\mathbf{Q}, E)$ were analyzed in the incoherent-scattering approximation after substracting background, multiphonon and multiple scattering as well as the elastic peak using the DAVE[61] and GetDOS programs.[62] Integration of the $S(\mathbf{Q}, E)$ spectra over the range of momentum transfers $0.4 \leq \mathbf{Q} \leq 6.0$ Å$^{-1}$ leads to neutron-weighted phonon DOSs for the samples.

**First-principles calculations:** The *ab initio* calculations were performed by using the projector augmented wave (PAW) method, as implemented in the Vienna *ab initio* simulation package (VASP).[63, 64] The generalized gradient approximation (GGA) was used for the exchange–correlation functional,[65] and a plane-wave energy cutoff of 450 eV was adopted. The phonon DOS and dispersions of ZrNiSn and Sb-doped ZrNiSn were calculated by using the frozen phonon method, as implemented in the Phonopy package.[66] The phonon group velocities, the scattering rates, and lattice thermal conductivities were obtained via the shengBTE package, in which we considered the interactions between atoms to their fourth nearest neighbors.[67] A 4*4*4 supercell (192 atoms) was used in both procedures. For the structural optimization of the unit cell, the k-points was chosen as 8*8*8, and the convergence accuracy of force was $10^{-5}$ eV/Å. For the supercells with displacements, only Γ point was considered. The energy convergence criterion was $10^{-7}$ eV throughout the work.

**Physical model of LO-TO splitting and screening effect:** The frequency (energy) difference between LO and TO arises from the long-range electric field generated by ionic polarization with the LO phonon vibration. Following the physical model proposed by Born and Huang,[51] the relation between LO and TO is described as:[68]

$$\omega_{\text{LO}}^2 = \omega_{\text{TO}}^2 + \Psi(\mathbf{q})\frac{|\mathbf{q}|^2}{\Omega}\left(\sum_a \frac{\mathbf{e_q} \cdot \mathcal{Z}_a \cdot \mathbf{e}_{\text{LO}}^a}{\sqrt{M_a}}\right)^2 \qquad (2)$$

where $\omega_{\text{LO}}$ and $\omega_{\text{TO}}$ are the frequencies of LO and TO, $\mathbf{e_q}$ is the unit vector along the momentum direction of $\mathbf{q}$, $\mathbf{e}_{\text{LO}}^a$ is the $|\mathbf{q}| \to 0$ limit of the eigenvector of the dynamical matrix of the LO mode, $M_a$ is the mass of atom $a$ in unit cell, $\mathcal{Z}_a$ is the Born effective charge of atom $a$, which reflect the change in polarization in response to the atom displacement, and $\Psi(\mathbf{q})$ is the Coulomb interaction in Fourier space, defined as:

$$\Psi(\mathbf{q}) = \frac{e^2}{\varepsilon_\infty |\mathbf{q}|^2} \qquad (3)$$

where $\varepsilon_\infty$ is the high-frequency (optical) dielectric function. In the limit of $|\mathbf{q}| \to 0$, the equation (1) could be simply replaced by the Lyddane-Sachs-Teller relationship: $\omega_{\text{LO}}^2/\omega_{\text{TO}}^2 = \varepsilon_s/\varepsilon_\infty$, $\varepsilon_s$ is the static dielectric function.[69]

Presence of free carriers and the concomitant screening effect can modify both the electron-LO phonon interaction and the dispersion characteristics of the LO optical phonons. Then, the Coulomb interaction in the momentum space is modified by a factor of $|\mathbf{q}|^2 r_{TF}^2/(1 + |\mathbf{q}|^2 r_{TF}^2)$ according to the Thomas-Fermi approximation, and then the equation (3) can be rewritten as:

$$\Psi(\mathbf{q}) = \frac{e^2 r_{TF}^2}{\varepsilon_\infty (1 + |\mathbf{q}|^2 r_{TF}^2)} \qquad (4)$$

Where $r_{\text{TF}}$ is the Thomas-Fermi screening length (or Debye length), and is defined as:

$$r_{\text{TF}} = \sqrt{\frac{\varepsilon k_B T}{n e^2}} \qquad (5)$$

Where $k_B$ is the Boltzmann constant. Higher $n$ leads to smaller $r_{\text{TF}}$, and smaller $r_{\text{TF}}$ means better screening effect. Enhanced screening effect leads to a smaller modified Coulomb interaction, and hence smaller LO-TO splitting as schematized in Figure 4(c).

It is necessary to point out that the screening model used here is in a static (Thomas-Fermi) approximation. It assumes that the scattering potential is static, or charge carriers are fast enough in response to movement of scattering potential, so that a screening scenario could form. This assumption is valid for static ionized impurity and grain boundary scatterings. However, it become more complex to deal with the polar optical phonon scattering. If the plasma (collective oscillation of free charges) frequency, $\omega_P = \sqrt{ne^2/\varepsilon_\infty m^*}$, is smaller than the optical phonon frequency $\omega_{\text{op}}$, the charge carriers lag behind the lattice vibration, which produces an anti-screening effect, and enhances the LO-TO splitting and the polar optical phonon scattering.[58, 70] Only when $\omega_P > \omega$, can the charge carriers instantaneously respond to the lattice vibration, and then meets the prerequisite for the approximation of static screening.[58, 70] In the case of ZrNiSn-based half-Heusler compounds, $\omega_P$ is estimated as 64.28 THz when $n = 1 \times 10^{20}$ cm$^{-3}$ (the high frequency dielectric parameter was calculated as $\varepsilon_\infty = 22.07$ and the calculated effective mass of electron $m^* = 2.8\pm0.2\ m_e$ are used here[40]). This value of $\omega_P$ is much larger than frequency of the longitudinal optical phonon, ~7.25 THz (or ~30 meV), and the Thomas-Fermi screening model is valid in the analysis of the screening of polar optical phonon scattering here.

## Data availability

The data that support the findings of this study are available from the corresponding authors on reasonable request.

**Figures**

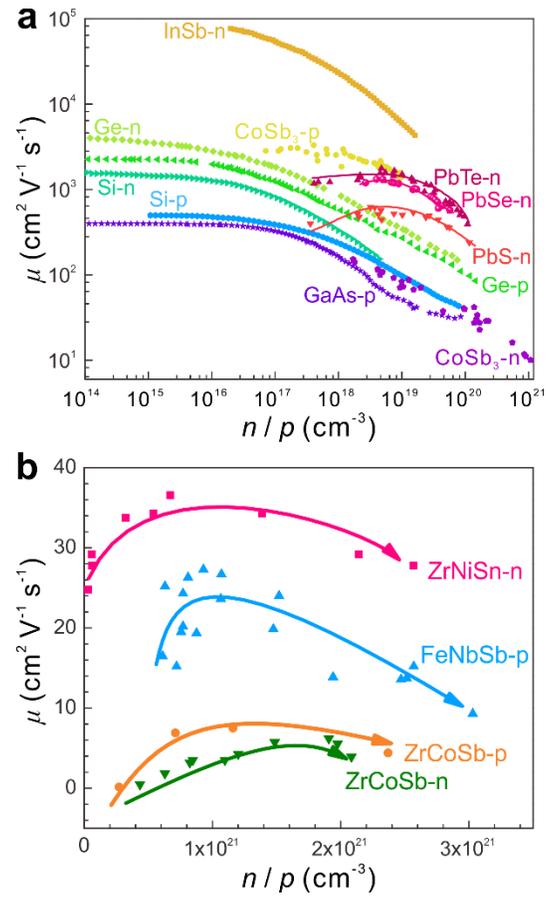

**Figure 1. Room temperature carrier mobility versus carrier concentration $n\,/\,p$.** (a) Some selected TE semiconductors,[15, 28, 29, 30, 31] and (b) several half-Heusler TE compounds.[40, 41, 42, 43]

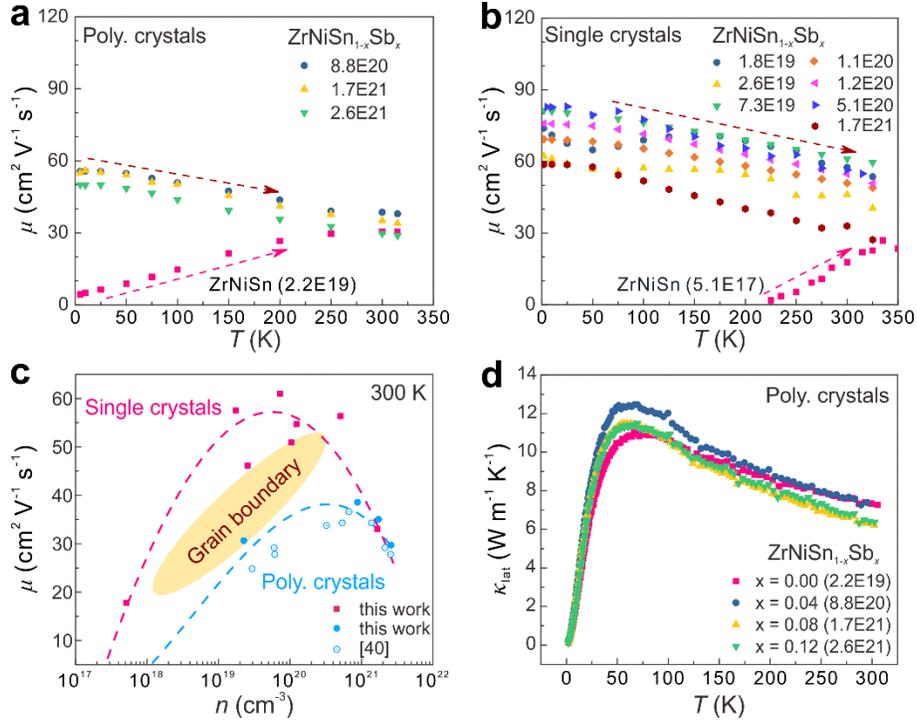

**Figure 2. Electrical and thermal transport properties for the parent and Sb-doped ZrNiSn$_{1-x}$Sb$_x$.** Carrier concentration dependent mobility for (a) polycrystalline samples and (b) single crystals. (c) Carrier mobility versus carrier concentration for ZrNiSn$_{1-x}$Sb$_x$ at 300 K. The dash lines in the figures are guides of the eye. (d) Lattice thermal conductivity, $\kappa_{lat}$, of the polycrystalline samples over the temperature range of 2 K to 300 K.

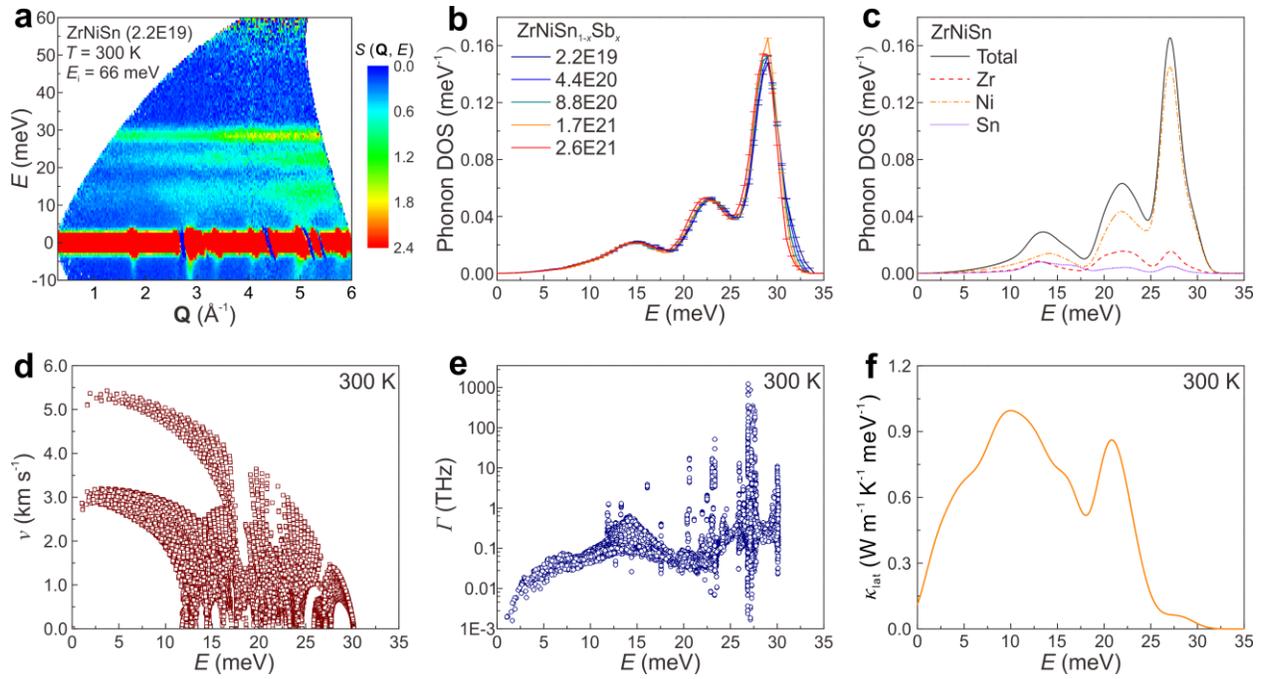

**Figure 3. Phonon DOSs and calculated lattice thermal conductivity.** (a) Experimental dynamical structure factor, $S(\mathbf{Q}, E)$, measured with inelastic neutron scattering for the ZrNiSn sample with $E_i$ = 66 meV. (b) Neutron-weighted phonon DOS of the ZrNiSn$_{1-x}$Sb$_x$ compounds with different carrier concentration $n$ (cm$^{-3}$). (c) Total and partial neutron-weighted phonon DOS for ZrNiSn obtained from first-principles calculations. (d) Calculated group velocities, $v$, and (e) scattering rate, $\Gamma$, for ZrNiSn. (f) calculated energy-dependent lattice thermal conductivity for ZrNiSn at 300 K.

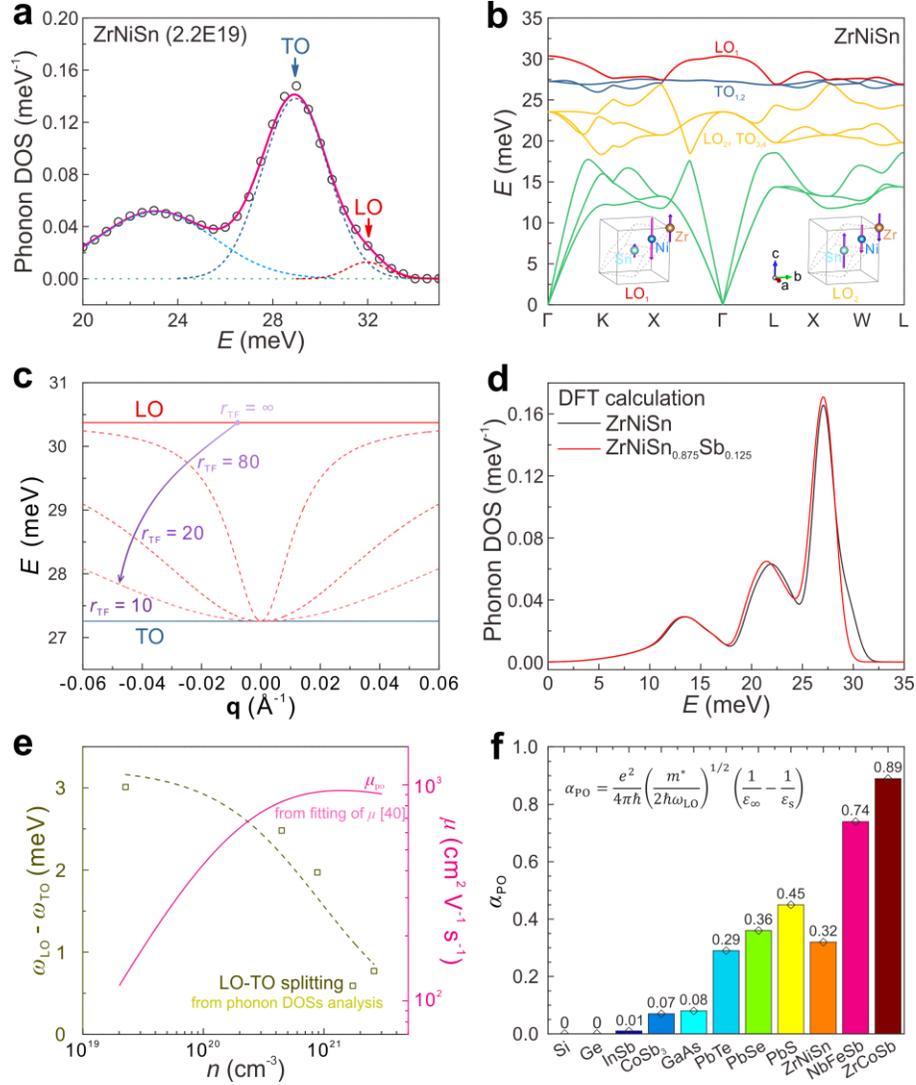

**Figure 4. Screening of the LO-TO splitting and polar optical scattering.** (a) Decomposition of the optical phonon bands shows the LO-TO splitting. (b) Calculated phonon dispersion in ZrNiSn parent sample. The insets illustrate the phonon vibration modes for the two longitudinal optical phonons (a propagation vector along $c$ axis is used here to define longitudinal and transverse phonon vibrations). The LO-TO splitting mainly occurs for the high frequency optical branches of $LO_1$ and $TO_{1,2}$. (c) Schematic illustration of the screening effect on the LO-TO splitting (following the Equations (2) and (4) in Methods) with different Thomas-Fermi screening lengths, $r_{TF}$ (in units of the lattice constant). With enhanced screening or smaller $r_{TF}$ as traced by the arrow, the LO frequency around the Brillouin zone center is suppressed. The Born effective charge, $\mathcal{Z}_a$, is supposed to be insensitive to the presence of free carriers, for this purpose, which is an approximation. (d) Comparison of the neutron-weighted phonon DOSs between ZrNiSn and ZrNiSn$_{0.875}$Sb$_{0.125}$ calculated with first principles. (e) Variation of the LO-TO splitting (or polarization field) with $n$. The dash curve is fitted using the Equations (2) and (4) in Methods including screening. The solid curve is the $n$-dependent mobility limited by polar optical phonon scattering in ZrNiSn$_{1-x}$Sb$_x$ compounds.[40] With increasing screening, the LO-TO splitting and polarization field are reduced, and the polar optical phonon scattering is reduced. (f) The polar coupling constant for some selected TE semiconductors as listed in Figure 1. The value for the half-Heusler compounds is calculated

based on the parameters as shown in Supplementary Table 1, while the data for other compounds is from literature.[27]

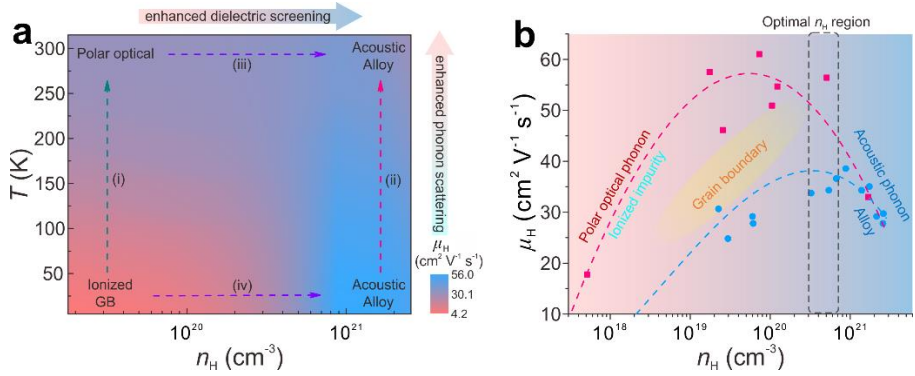

**Figure 5 Carrier scattering phase diagram for ZrNiSn-based half-Heusler compounds.** (a) The phase diagram as functions of temperature and carrier concentration. GB represent grain boundary scattering. (b) The phase diagram with single-crystalline and polycrystalline data as function of carrier concentration at 300 K. The dash lines in the figures are guides of the eye.

## Acknowledgements

J.M. thanks the financial support from the National Science Foundation of China (No. 11774223 and No. U1732154) and a Shanghai talent program. C.G.F. thanks the financial support from the Deutsche Forschungsgemeinschaft (DFG, German Research Foundation) – Projektnummer (No. 392228380) and the Alexander von Humboldt Foundation. J.Y. acknowledges the financial support from Natural Science Foundation of China (No. 11674211 and No. 51761135127), and the 111 project D16002. T.J.Z. thanks the financial support from the National Science Foundation of China (No. 51761135127) and the National Science Fund for Distinguished Young Scholars (No. 51725102). Q.Y.R. thanks Dr. Jiao Lin from the Oak Ridge National Lab for the help on the phonon DOSs analysis using the GetDOS programs, and Dr. Xiaokun Gu for the discussion about LO-TO splitting. The inelastic neutron scattering data using HRC were supported with the general proposal (Proposal No. 2017A0071 & 2018B0281) of J-PARC/MLF. Some of the neutron diffraction data using SuperHRPD were taken under both the general proposal (Proposal No. 2017A0118) and the fast track proposal (Proposal No. 2017BF0804) of J-PARC/MLF, while some data using GPPD were taken under the proposal (Proposal No. P2018091800008) of CSNS.

## Author contributions

Q.Y.R., Z.Y.L., T.M., S.A. and J.M. designed and performed the INS experiment using HRC at the J-PARC. Q.Y.R., M.H., S.H.L., S.T., T.K. and J.M. designed and performed the high-resolution neutron diffraction experiment using superHRPD at the J-PARC. Q.Y.R., L.H.H., X. T. and J. M. designed and performed the neutron diffraction experiment using GPPD at the CSNS. C.G.F. and Q.Y.Q. synthesized the samples. C.G.F. and C.F. performed the transport properties measurement. Q.Y.R. analyzed the experiment data, along with all co-authors. J.Y. and S.N.D. performed the First-principles calculations. Q.Y.R., J.Y., D.J.S., and J.M. interpreted the LO-TO splitting phenomenon. Q.Y.R., C.G.F., T.J.Z., J.Y. and J.M. convinced the project and drafted the manuscript. All authors analyzed and reviewed the results, and provided input to this paper.

## Competing interests

The authors declare no competing interests.

# Supplement Information

# Establishing the carrier scattering phase diagram for ZrNiSn-based half-Heusler thermoelectric materials

Qingyong Ren[1], Chenguang Fu[2], Qinyi Qiu[3], Shengnan Dai[4], Zheyuan Liu[1], Takatsugu Masuda[5], Shinichiro Asai[5], Masato Hagihala[6], Sanghyun Lee[6], Shuki Torri[6], Takashi Kamiyama[6,7], Lunhua He[8,9,10], Xin Tong[10,11], Claudia Felser[2], David J. Singh[12], Tiejun Zhu[3], Jiong Yang[4], and Jie Ma[1,13]

[1]Key Laboratory of Artificial Structures and Quantum Control, School of Physics and Astronomy, Shanghai Jiao Tong University, 800 Dongchuan Road, Shanghai 200240, China
[2]Max Planck Institute for Chemical Physics of Solids, Nöthnitzer Straße 40, 01187 Dresden, Germany
[3]State Key Laboratory of Silicon Materials, School of Materials Science and Engineering, Zhejiang University, Hangzhou 310027, China
[4]Materials Genome Institute, Shanghai University, 99 Shangda Road, Shanghai 200444, China
[5]Neutron Science Laboratory, Institute for Solid State Physics, University of Tokyo, Kashiwanoha, Kashiwa, 277-8581, Japan
[6]Institute of Materials Structure Science, High Energy Accelerator Research Organization (KEK), Tokai, Ibaraki 319-1106, Japan
[7]Department of Materials Structure Science, Sokendai (The Graduate University for Advanced Studies), Tokai, Ibaraki 319-1106, Japan
[8]Beijing National Laboratory for Condensed Matter Physics, Institute of Physics, Chinese Academy of Sciences, Beijing 100190, China
[9]Songshan Lake Materials Laboratory, Dongguan, Guangdong 523808, China
[10]Spallation Neutron Source Science Center, Dongguan 523803, China
[11]Institute of High Energy Physics, Chinese Academy of Sciences, Beijing 100049, China
[12]Department of Chemistry and Department of Physics and Astronomy, University of Missouri-Columbia, Columbia, MO 65211, USA
[13]Shenyang National Laboratory for Materials Science, Institute of Metal Research, Chinese Academy of Sciences, Shenyang, 110016, China

Correspondence and requests for materials should be addressed to C.G.F. (email: Chenguang.Fu@cpfs.mpg.de) or to J.Y. (email: jiongy@t.shu.edu.cn) or to J.M. (email: jma3@sjtu.edu.cn).

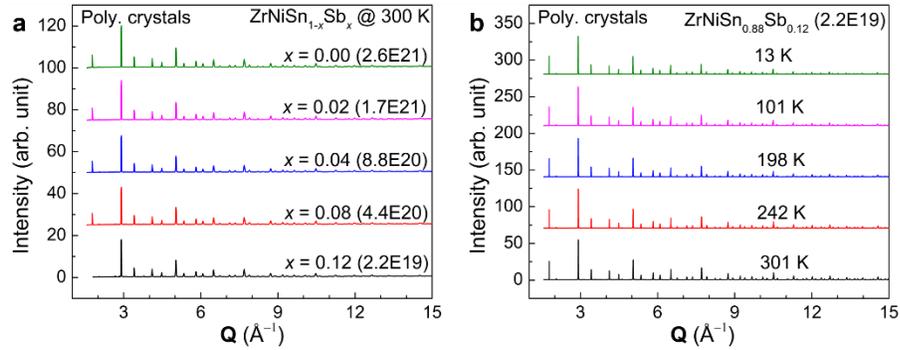

**Supplementary Figure 1 Neutron powder diffraction patterns** of (a) polycrystalline ZrNiSn$_{1-x}$Sb$_x$ with different carrier concentrations collected at 300 K on the GPPD, CSNS and of (b) ZrNiSn$_{0.88}$Sb$_{0.12}$ collected at different temperatures on the SuperHRPD, J-PARC. Rietveld refinements show that all the polycrystalline samples contain 5-7 % more Ni at the *4d* (¾, ¾, ¾) vacancy position, being consistent with the EPMA (electron probe microanalysis) analysis.[40]

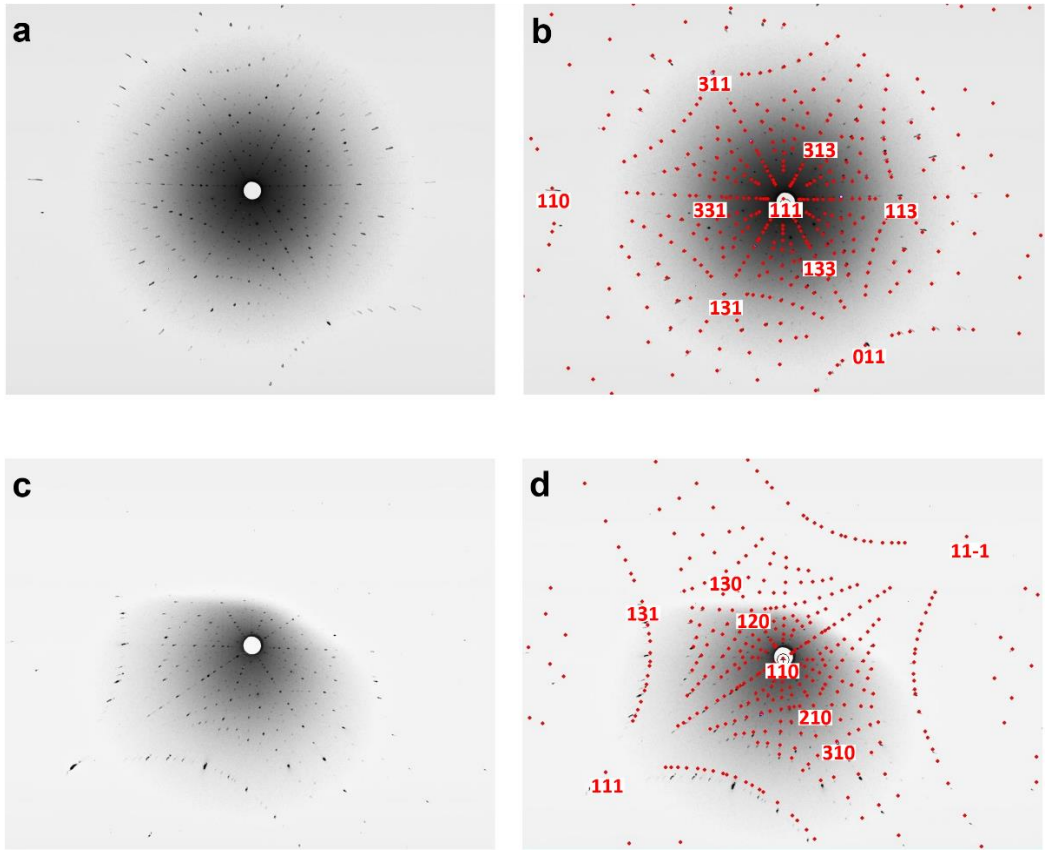

**Supplementary Figure 2 Orientation of the single crystals with Laue diffraction.** (a) The diffraction pattern for a typical (111) surface with the simulated pattern and Miller indices in (b). (c) The diffraction pattern for a typical (110) surface with the simulated pattern and Miller indices in (d).

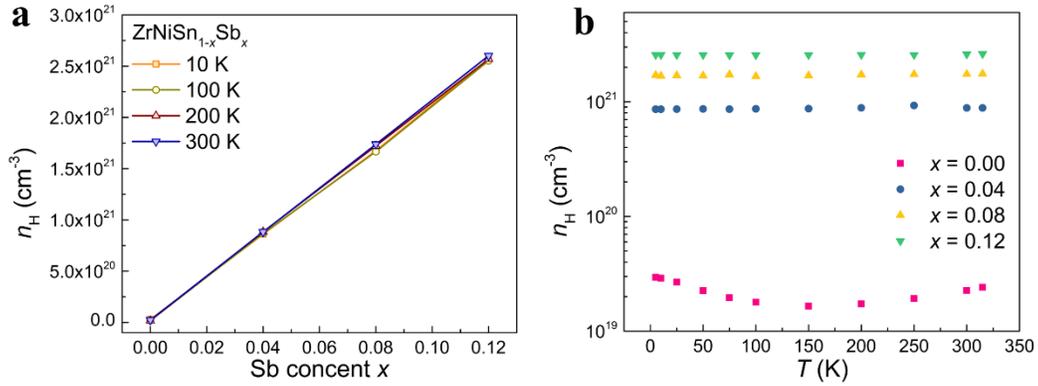

**Supplementary Figure 3 Hall carrier concentration $n_H$.** (a) $n_H$ as a function of Sb content at 10 K, 100 K, 200 K and 300 K, (b) $n_H$ as a function of temperature for ZrNiSn$_{1-x}$Sb$_x$ over the temperature range of 5 K to 315 K. The almost T-independent $n_H$ indicates that the ions should be fully ionized below 5 K. This excludes the effect of increased $n_H$ on the discussion of carrier mobility in Figure 2.

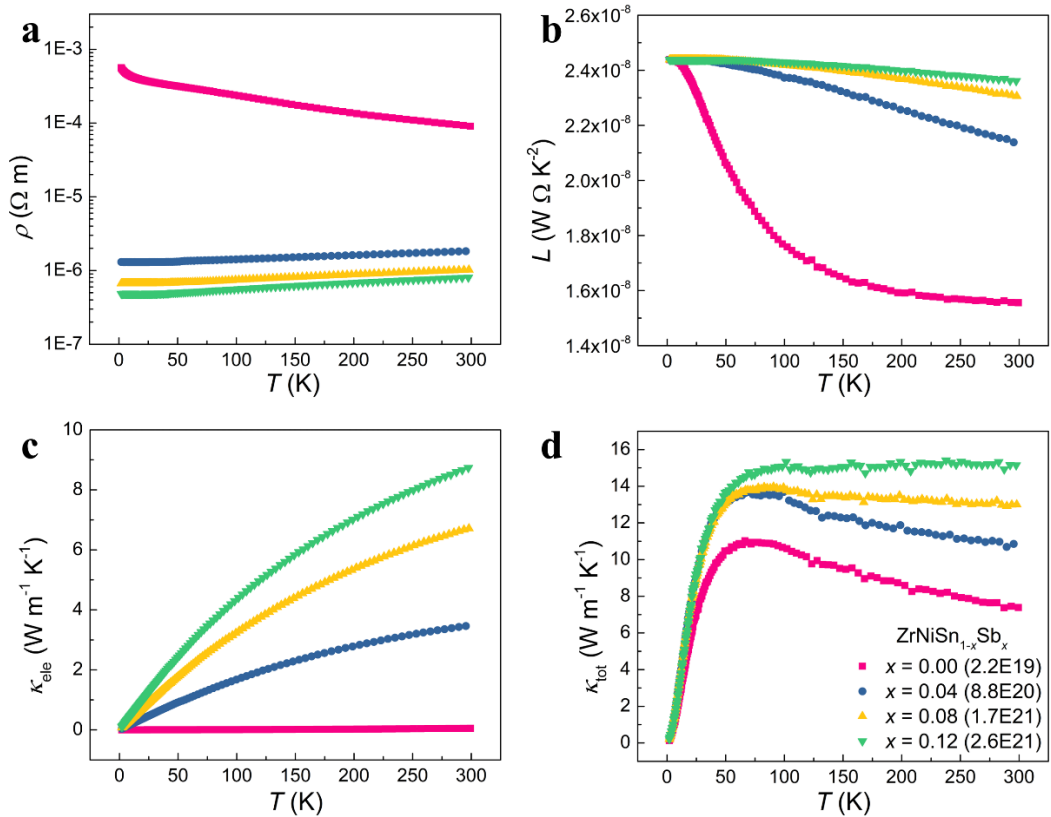

**Supplementary Figure 4 Transport properties as functions of temperature.** (a) Electronic resistivity $\rho$, (b) Lorenz number $L$, (c) calculated electronic contribution to thermal conductivity $\kappa_{ele}$, (d) measured total thermal conductivity $\kappa_{tot}$. The $\kappa_{ele}$ is estimated with the Wiedemann–Franz law, $\kappa_{ele} = LT/\rho$.

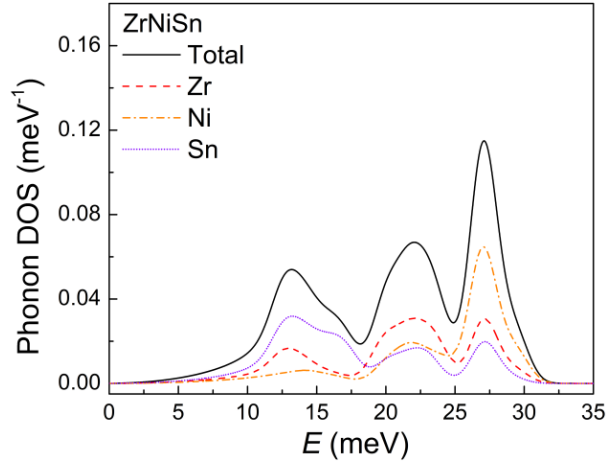

**Supplementary Figure 5 Calculated phonon density of states (DOSs).** (a) Total and partial phonon DOS, without considering difference of atomic mass, $m$, and neutron scattering cross section, $\sigma$, between different elements, for ZrNiSn obtained from first-principles calculations. The neutron-weighted phonon DOS for ZrNiSn are shown in Figure 3(c) in the main context. In inelastic neutron scattering measurements for ZrNiSn$_{1-x}$Sb$_x$ compounds, the contribution from Ni element is overemphasized: $g_{\text{NW}}(E) = \left[\frac{\sigma_{\text{Zr}}}{m_{\text{Zr}}} g_{\text{Zr}}(E) + \frac{\sigma_{\text{Ni}}}{m_{\text{Ni}}} g_{\text{Ni}}(E) + (1-x)\frac{\sigma_{\text{Sn}}}{m_{\text{Sn}}} g_{\text{Sn}}(E) + x\frac{\sigma_{\text{Sb}}}{m_{\text{Sb}}} g_{\text{Sb}}(E)\right]/3$, where $g_{\text{Zr}}(E)$, $g_{\text{Ni}}(E)$, $g_{\text{Sn}}(E)$ and $g_{\text{Sb}}(E)$ are the partial densities of states of Zr, Ni, Sn and Sb, respectively. The values of $\sigma/m$ for (Zr, Ni, Sn, Sb) are (0.07081, 0.31520, 0.03471, 0.03203) b/amu, respectively.

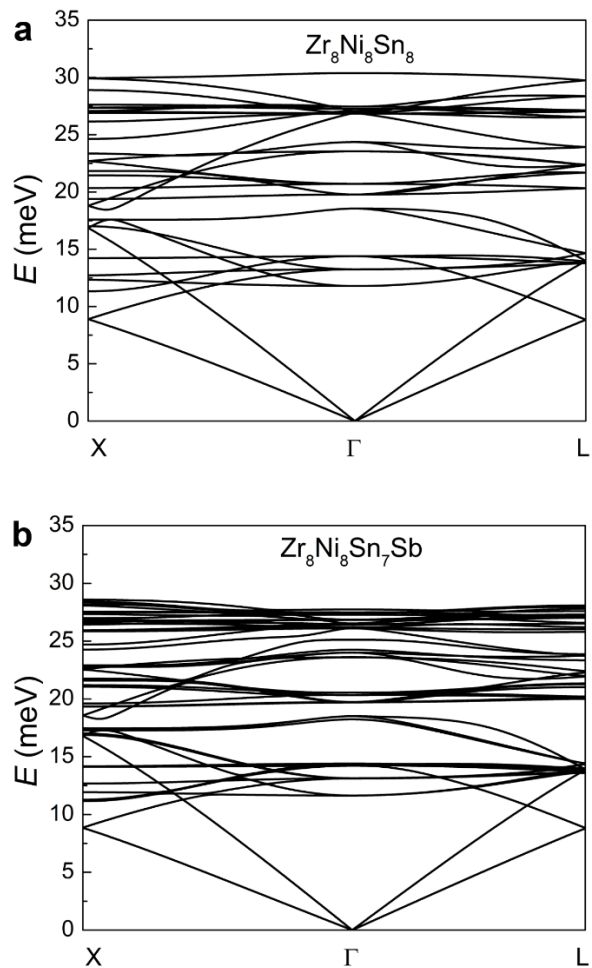

**Supplementary Figure 6 First-principles calculation of the phonon dispersions** in the supercell (a) ZrNiSn (or $Zr_8Ni_8Sn_8$) and (b) $ZrNiSn_{0.875}Sb_{0.125}$ (or $Zr_8Ni_8Sn_7Sb$). In contrast to the obvious LO-TO splitting in $Zr_8Ni_8Sn_8$, doping of Sb on Sn site breakdown this splitting due to increased screening effect.

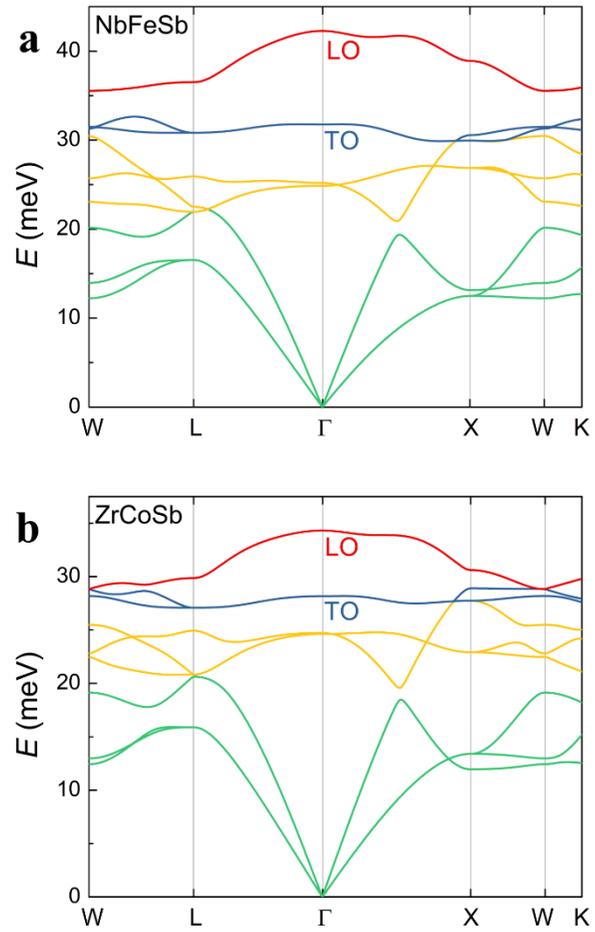

**Supplementary Figure 7 First-principle calculation of the phonon dispersions** for (a) NbFeSb and (b) ZrCoSb. Both samples exhibit obvious LO-TO splitting.

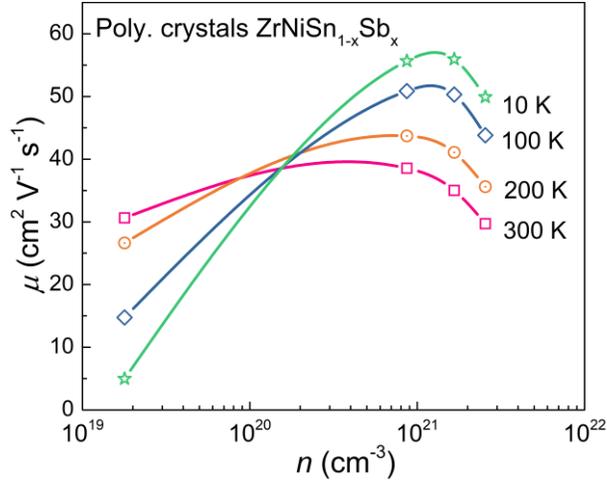

**Supplementary Figure 8 Carrier mobility as a function of carrier concentration $n$** in polycrystalline ZrNiSn$_{1-x}$Sb$_x$ samples at 10 K, 100 K, 200 K and 300 K. The solid lines are guides of the eye.

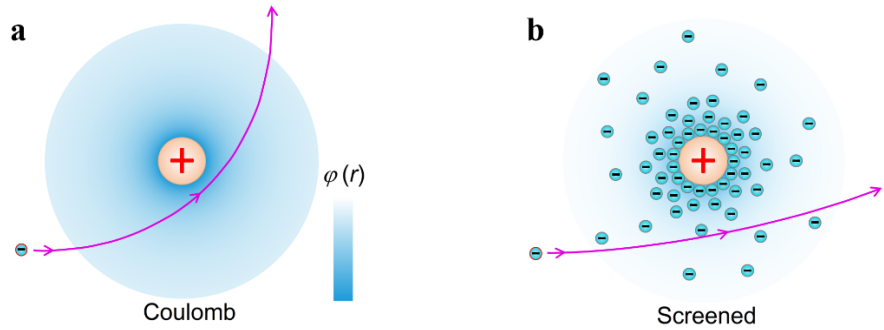

**Supplementary Figure 9 Schematic demonstration of the screened ionized impurity scattering.** (a) Bare Coulomb potential of an ionized impurity center ($\varphi(r) \propto \frac{1}{r}$) as a function of distance $r$ in real space and (b) screened potentials ($\varphi(r) \propto \frac{1}{r}\exp(-\frac{r}{r_{TF}})$) by free carriers in the approximation of Thomas-Fermi model. Compared with the bare Coulomb potential, the screened potential drops more quickly with distance, and the momentums of traveling charge carriers are less changed when they pass through the ionized impurity scattering centers.

**Supplementary Table 1 The polar coupling constant for ZrNiSn, NbFeSb and ZrCoSb.**

| Material | $\omega_{LO}/\omega_{TO}$ | $\varepsilon_\infty$ | $\varepsilon_s$ | $\hbar\omega_{LO}$ | $m^*$ ($m_e$) | $\alpha_{PO}$ |
|---|---|---|---|---|---|---|
| ZrNiSn | 1.243 | 22.072 | 27.435 | 30.328 | 2.8 | 0.32 |
| NbFeSb | 1.771 | 27.475 | 48.659 | 42.273 | 6.9 | 0.74 |
| ZrCoSb | 1.485 | 18.628 | 27.661 | 34.312 | 6.5 | 0.89 |

The parameters, $\omega_{LO}$, $\omega_{TO}$, $\varepsilon_\infty$, $\varepsilon_s$ and $\hbar\omega_{LO}$, are determined from the first-principle calculations, and $m^*$ are cited elsewhere.[33, 40, 42] The phonon dispersions for NbFeSb and ZrCoSb are shown in Supplementary Figure 7, which show obvious LO-TO splitting.